\documentclass[12pt]{article}

\newcommand{\add}{\addtocounter{eqncnt}{1}}
\newcounter{eqncnt}[section]

\newcommand{\be}{\begin{equation}}
\newcommand{\ee}{\end{equation}\add}
\newcommand{\bea}{\begin{eqnarray}}
\newcommand{\eea}{\end{eqnarray}}

\newcommand{\bra}{\left\langle}
\newcommand{\ket}{\right\rangle}

\begin{document}
\begin{center}
{\Large \bf Averaging in cosmological models using scalars} \\[2mm]

\vskip .5in

{\sc A.A. Coley}\\
{\it Department of Mathematics and Statistics}\\
{\it Dalhousie University, Halifax, NS B3H 3J5, Canada}\\
{\it aac@mathstat.dal.ca }
\end{center}

\begin{abstract}

The averaging problem in cosmology is of considerable importance
for the correct interpretation of cosmological data. A rigorous
mathematical definition of averaging in a cosmological model is
necessary. In general, a spacetime is completely characterized by
its scalar curvature invariants, and this suggests a particular
spacetime averaging scheme based entirely on scalars. We clearly
identify the problems of averaging in a cosmological model. We
then present a precise definition of a cosmological model, and
based upon this definition, we propose an averaging scheme in
terms of scalar curvature invariants. This scheme is illustrated
in a simple static spherically symmetric perfect fluid
cosmological spacetime, where the averaging scales are clearly
identified.

\end{abstract}



\newpage

\section{The averaging problem in cosmology}

Cosmological observations \cite{Riess:2006fw,bennett}, based on
the assumption of a spatially homogeneous and isotropic
Friedmann-Lema\^{i}tre-Robertson-Walker (FLRW) model plus small
perturbations, are usually interpreted as implying that there
exists dark energy, the spatial geometry is flat, and that there
is currently an accelerated expansion giving rise to the so-called
$\Lambda$CDM-concordance model. Although the concordance model is
quite remarkable, it does not convincingly fit all data
\cite{refs}. Unfortunately, if the underlying cosmological model
is not a perturbation of an exact flat FLRW solution, the
conventional data analysis and their interpretation is not
necessarily valid. For example, the standard analysis of type Ia
supernovae (SNIa) and CMB data in FLRW models cannot be applied
directly when backreaction effects are present, because of the
different dynamical behaviour of the spatial curvature
\cite{Shapiro:2005}. Indeed, supernovae data can be explained
without dark energy in inhomogeneous models, where the full
effects of general relativity (GR) come into play. For example, it
has been indeed shown that the Lema\^{i}tre-Tolman-Bondi (LTB)
solution can be used to fit the observed data without the need of
dark energy \cite{LTBgeo}, although it is necessary to place the
observer at the center of a rather large-scale underdensity.
Therefore, the averaging problem in cosmology is of considerable
importance for the correct interpretation of cosmological data.
The correct governing equations on cosmological scales are
obtained by averaging the Einstein field equations (EFE) of GR
(plus a theory of photon propagation; i.e., information on what
trajectories actual particles follow). By assuming spatial
homogeneity and isotropy on the largest scales, the
inhomogeneities affect the dynamics though correction
(backreaction) terms, which can lead to behaviour qualitatively
and quantitatively different from the FLRW models.

\subsection{General Approaches}

The gravitational field equations on large scales are obtained by
averaging the EFE of GR. It is
necessary to use an exact covariant approach which gives a
prescription for the correlation functions that emerge in an
averaging of the full tensorial EFE. The Universe is not isotropic
or spatially homogeneous on local scales. An averaging of
inhomogeneous spacetimes on large scales can lead to important
effects. For example, on cosmological scales the dynamical
behavior can differ from that in the spatially homogeneous and
isotropic FLRW model
\cite{aver}; in particular, the expansion rate may be
significantly affected. Consequently, a solution of the averaging
problem is of considerable importance for the correct
interpretation of cosmological data.  The averaging problem in GR and
cosmology is of fundamental
importance.

There are a number of approaches to the averaging problem
\cite{aver,Zala,buch}. In the approach of Buchert \cite{buch}  a
3+1 cosmological space-time splitting is employed and only scalar
quantities are averaged. The perturbative approach involves
averaging the perturbed Einstein equations; however, a
perturbation analysis cannot provide any information about an
averaged geometry. In the space-time or space volume averaging
approach tensors, and in some cases only scalar quantities, are
averaged; this procedure is not generally covariant hence the
results are somewhat limited and the conclusions unreliable. In
all of these approaches, in analogy with Lorentz's approach to
electrodynamics,  an averaging of the Einstein equations is
performed to obtain the averaged field equations.

However, to date the macroscopic gravity (MG)  approach is the
only approach to the averaging problem in GR \cite{Zala} which
gives a prescription for the correlation functions which emerge in
an averaging of the non-linear field equations (without which the
averaging of the Einstein equations simply amount to definitions
of the new averaged terms). The MG approach is a fully covariant,
gauge independent and exact method. The space-time averaging
procedure adopted in MG for any differentiable
manifold is based on the concept of Lie-dragging of
averaging regions, and it has been proven to exist
on an arbitrary Riemannian space-times with well-defined local
averaged properties. Averaging of the structure equations for the
geometry of GR brings about the structure equations for the
averaged (macroscopic) geometry and the definitions and the
properties of the correlation tensors. The averaged Einstein
equations can always be written in the form of the Einstein
equations for the macroscopic metric tensor when the correlation
terms are moved to the right-hand side of the averaged Einstein
equations to serve as the geometric modification to the averaged
(macroscopic) matter energy-momentum tensor \cite{Zala}.

The formal mathematical issues of averaging tensors on a
differential manifold has recently been revisted \cite{newaver}.
However, we note that integrating scalars on spacetime regions is
always well-defined.

\subsection{Scales}

In any theory of physics, the scales over which the physical
theory is applicable must be specified \cite{Ellis}. There is a
hierarchy of cosmological scales of physical interest.
Consequently, we must specify the cosmological scale over which
averaging occurs (i.e., we must specify the averaging scale $\ell$
or averaging region, which then determines the type of averaging
or smoothing that occurs). In particular, a given geometry may be
inhomogeneous on both small scales $\ell_s$  (i.e., local scales
such as local density inhomogeneities) and large scales $\ell_l$
with $\ell_s << \ell_l$ (where $\ell_l \sim \ell_H$, and $\ell_H$
is the Hubble scale). The averaging scale $\ell$ will satisfy
$\ell_s << \ell$, but it is possible that $\ell << \ell_l$, so
that the geometry is still inhomogeneous (but `smooth') on large
scales. In cosmology it is assumed that the averaging scale $\ell$
is bigger than scale of largest observed structures (clusters of
galaxies) and voids and that $\ell < \ell_H$ \cite{BuchCar}. There
is also the homogeneity scale, $\ell_{hom}$, the largest scale any
inhomogeneities are observed. It is usually assumed that
$\ell_{hom}<\ell$ $(< \ell_H)$.

The physical description of a cosmological model depends on the
averaging scale \cite{BuchEll}. The scales $\ell_s$, $\ell_l$,
$\ell$ in the cosmological model and the range of validity (i.e.,
$\ell_s << \ell < \ell_l$) must be specified. Of course, the range
of scales relevant to cosmology are the largest scales of
averaging, larger than the largest scale of cosmological
structures  and comparable to a fraction of $\ell_H$. In the
context of perturbation theory, particular attention is paid to
both the scale of inhomogeneities in the background and the scale
of inhomogeneities of the perturbations \cite{BuchEllElst}.

\subsection{Spatial curvature}

In \cite{CPZ} the MG equations were explicitly solved in a FLRW
background geometry and it was found that the correlation tensor
(backreaction) is of the form of a spatial curvature. Thus, the
averaged Einstein equations for a flat spatially homogeneous,
isotropic macroscopic space-time geometry has the form of the
Einstein equations of GR for a non-flat spatially homogeneous,
isotropic space-time geometry.

The relevance of spatial curvature in realistic models of the
universe that describe the dynamics of structure formation since
the epoch of last scattering was discussed in \cite{BuchEllElst};
in particular, in arguments about spatial curvature in
perturbation theory, the quasi-Newtonian approximation must be
used with care since spatial curvature is an inherently
relativistic phenomenon (that does not occur in Newtonian
physics). We note that a spatially dependent constant spatial
curvature $k$ can alleviate the tension in observational data
\cite{Col}. Indeed, if the spatial curvature parameter $k$ is
allowed to be a function of position, then considerable spatial
curvature (locally) is permissable (consistent with CMB
observations) \cite{Clifton}.

\subsection{Discussion}

Clearly, backreaction (averaging) effects are real, but their
relative importance must be determined \cite{coleynull}.
Observational data suggests a normalized spatial curvature
$|\Omega_{k}| \approx 0.01 - 0.02$ (i.e., of about a percent).
Combining these observations with large scale structure
observations then puts stringent limits on the curvature parameter
in the context of adiabatic $\Lambda$CDM models; however, these
data analyses are very model- and prior-dependent
\cite{Shapiro:2005}, and care is needed in the proper
interpretation of the data. There is a heuristic argument that
$\Omega_{k}\sim 10^{-3}-10^{-2}$ \cite{NewRas,Col}, which is
consistent with CMB observations
\cite{Riess:2006fw,bennett,refs,Shapiro:2005}  and agrees with
estimates for intrinsic curvature fluctuations using realistically
modelled clusters and voids in a Swiss-cheese model. It must be
appreciated that such a value for $\Omega_{k}$, at the 1\% level,
is relatively large and may have a significant dynamical effect on
the evolution of the universe and the interpretation of
cosmological observations.

Note that in a scenario in which $|\Omega_{k}|\sim 0.01-0.02$, the
current contribution from the spatial curvature is much greater
than the energy density of radiation, and is comparable to the
energy density in luminous matter. In addition, such a value
cannot be naturally explained by inflation. From standard
analysis, depending on the initial conditions and the details of a
specific model of inflation, $|\Omega - 1|$ would be extremely
small. Indeed, any value for $\Omega_{k}$ at the 1 \% level would
be very difficult to explain within the theory of inflation;
therefore, any non-zero residual curvature at this level can only
be naturally explained in terms of an averaging effect.

\subsection{Null geodesics}

Ultimately we wish to determine the effects of averaging or
backreaction on the evolution of the universe and the
interpretation of cosmological observations.
All deductions about cosmology are based on light paths. Only the
redshift and the energy flux of light arriving from a distant
source are observed, rather than the expansion rate or the matter
density. It is often assumed that intervening inhomogeneities
average out. However, inhomogeneities affect curved null geodesics
\cite{NewRas,buch} and can drastically alter observed distances
when they are a sizable fraction of the curvature radius. In the
real universe, voids occupy a much larger region as compared to
structures, hence light preferentially travels much more through
underdense regions and the effects of inhomogeneities on
luminosity distance are likely to be significant.

The effect of averaging null geodesics in inhomogeneous models was
discussed in \cite{coleynull}.  GR is treated as a microscopic
(classical) theory. Real photons travel on null geodesics in the
microscopic geometry. However, because all observations are of
finite resolution, observations necessarily involve averages of
measured quantities. Therefore, in interpreting real observations,
it is necessary to model properties of (not only a single photon
but of) a `narrow' beam or bundle of photons (i.e., a local
congruence of null geodesics). From the geometric optics
approximation we can obtain the optical scalar (Dyer-Roeder)
equations that govern the propagation of the local shearing and
expansion (of the cross-sectional area of the beam) with respect
to the affine parameter along the congruence due to Ricci
focussing and Weyl tidal focussing \cite{Dyer}. Since the {\em
nonlinear} optical scalar equations require integration along the
beam, the optics for a lumpy distribution does not average and
there may be important resulting effects. Similar issues have been
discussed recently from a different point of view
\cite{NewRas,marra-sc}. The formulation of the EFE on the null
cone was discussed in \cite{NEWCOLEY}. Clearly averaging can have
an important effect on photon propagation and hence observations.

In particular, assuming GR to be  a microscopic theory on small
scales with local metric field ${\bf g}$ (the micro-geometry) and
matter fields, a photon follows a null geodesic ${\bf k}$ in the
local geometry, and after averaging we obtain  \cite{coleynull} a
smoothed out macroscopic geometry (with macroscopic metric
$\langle {\bf g} \rangle$) and macroscopic matter fields, valid on
larger scales. But, in general, the ``averaged'' vector $\langle
{\bf k} \rangle$ need not be null, need not be geodesic (and even
if it is, need not be affinely parametrized) in the
macro-geometry.

\newpage

\section{Scalar curvature invariants and averaging}

In \cite{NEWINV}, it was shown that the class of four-dimensional (4D)
Lorentzian manifolds that cannot be completely characterized by
the scalar polynomial curvature invariants constructed from the
Riemann tensor and its covariant derivatives must be of Kundt
form. This implies that, in general, a spacetime is completely
characterized by its scalar curvature invariants and this suggests
a particular spacetime averaging scheme based entirely on scalars.
Let us first review the main mathematical background.

For any given spacetime $({\bf M,g})$ we define the set of all
scalar invariants
\begin{equation}
\mathcal{I}\equiv\{R,R_{\mu\nu}R^{\mu\nu},C_{\mu\nu\alpha\beta}C^{\mu\nu\alpha\beta},
R_{\mu\nu\alpha\beta;\gamma}R^{\mu\nu\alpha\beta;\gamma},
R_{\mu\nu\alpha\beta;\gamma\delta}R^{\mu\nu\alpha\beta;\gamma\delta},\dots\}
\,. \nonumber
\end{equation}
\noindent Consider a spacetime $({\bf M,g})$ with a set of
invariants $\mathcal{I}$. Then, if there does not exist a
continuous metric deformation of $g$ having the same set of
invariants as $g$, we will call the set of invariants
\emph{non-degenerate}. Furthermore, the spacetime metric, $g$, will
be called \emph{$\mathcal{I}$-non-degenerate}. This implies that
for a metric which is $\mathcal{I}$-non-degenerate the invariants
characterize the spacetime uniquely, at least locally, in the
space of (Lorentzian) metrics. This means that these metrics are
characterized by their curvature invariants and therefore we can
distinguish such metrics using their invariants.

It was proven, on a case-by-case (depending on the algebraic type;
characterized by their Petrov and Segre types or, equivalently, in
terms of their Ricci, Weyl (and Riemann) types \cite {coley}),
that a 4D spacetime is either $\mathcal{I}$-non-degenerate, which
implies that the spacetime metric is determined  locally by its
curvature invariants, or the metric is a Kundt metric. This is a
striking result because it tells us that metrics not determined by
their curvature invariants must be of Kundt form. These Kundt
metrics therefore correspond to degenerate metrics in the sense
that many such spacetimes can have identical invariants. The Kundt
class is defined by those metrics admitting a null vector $\ell$
that is geodesic, expansion-free, shear-free and twist-free
{\footnote{ We recall that in the Riemannian case a manifold is
always locally characterized by its scalar polynomial
invariants.}} {\footnote{ We note that this exceptional property
of the the degenerate Kundt metrics essentially follows from the
fact that they do not define a unique timelike curvature
operator.}}.

Let us also consider the `inverse' question: given a set of scalar
polynomial invariants, what can we say about the underlying
spacetime? In practice, it is somewhat tedious and a lengthy
ordeal to determine the spacetime from the set of invariants.
In 4D we can partially
characterize the Petrov type in terms of scalar curvature
invariants. In most circumstances we only need some partial
results or necessary conditions or we are dealing with special cases. 
For example, we found that if
$27J^2\neq I^3$, or if $27J^2= I^3$ but the differential
invariants $S_1\neq 0$ or $S_2\neq 0$, then the spacetime is
$\mathcal{I}$-non-degenerate. Having determined when a spacetime
is completely characterized by its scalar curvature invariants, it
is also of interest to determine the minimal set of such
invariants needed for this classification.  It is also of interest
to study when a spacetime can be explicitly {\em constructed} from
scalar curvature invariants.

Let us return to the question of averaging. We have noted that a
general spacetime is completely characterized by its scalar
curvature invariants. Since we know how to average scalar
quantities, we can average all of the scalar curvature invariants
that can then represent an averaged spacetime (with that set of
averaged scalar invariants). We shall return to this in Section 4.
However, we note that cosmological models belong to the set of
spacetimes completely characterized by their scalar curvature
invariants, suggesting that we can average a cosmological model
using scalar invariants. We must first define a cosmological model
rigorously, which we do next.

\newpage

\section{Cosmological models and averaging}

There are a number of technical problems with averaging of tensor
fields on a differentiable manifold ${\bf M}$. Clearly we need a
covariant approach. Is the average of the metric the metric? 
{\footnote{ Note that if the average of the metric is not the metric itself,
then the conformal structure of the macro- and micro- geometries are not the same 
and photons will not follow null geodesics in the macrogeometry in general.}}
There
is the question of averaging verses smoothing. We want to avoid
issues with respect to coordinates; i.e., if we choose more smooth
coordinates, we are not averaging but just using different
coordinates to represent the same geometry. It may be possible to
avoid several of these technical problems by adopting an
approach based  on scalar curvature invariants.  Note that
although such an approach may work for any differentiable manifold
which is \emph{$\mathcal{I}$-non-degenerate}, we shall focus on
the cosmological problem for the most part here.

Therefore, we wish to discuss the averaging problem in the context
of cosmology. Although this context may simplify the issue of
averaging in technical terms, there are some new problems of
principle that are then introduced. First, the cosmological model
is a {\em mixed} model, in that the matter is already assumed to
be averaged, but the geometry is not (necessarily) {\footnote{
Since the time dependence is smooth, perhaps some smoothing of the
microgeometry has already been assumed.}}. Therefore, we need a
consistent model for the matter, represented on the characteristic
averaging scale, and its appropriate (averaged) physical
properties {\footnote{ Note that a discrete (non-continuous fluid)
model for the cosmological matter has been discussed recently
\cite{CF}.}}. It is known that the separation between the
gravitational field and the matter is not scale invariant and the
notion of a perfect fluid is not scale invariant \cite{LL};
averaging (in the presence of a gravitational field) modifies the
equation of state of the matter \cite{Deb}. In addition, since
averaging does not conserve geodesics (in fact, averaging need not
even conserve the metric signature), we need further assumptions
in order to be able to compare the models with observational data.

First, a precise definition of a cosmological model is necessary;
i.e., a framework in which to do averaging. This definition
includes an appropriate way to do averaging and how to deal with
photons. In particular, it is necessary to make all of the
assumptions in the model clear.

The precise definition of a cosmological model we shall adopt,
based in part on \cite{Col,coleynull,NEWCOLEY}, is given by the
following conditions C1 -- C5 (cf. \cite{Ellis}):

\subsection{Definition of a cosmological model}

\subsection*{C1. Spacetime Geometry:} The spacetime geometry
$({\bf M,g})$  is defined by a smooth Lorentzian  metric ${\bf g}$
(characterizing the macroscopic gravitational field) defined on a
smooth differentiable manifold ${\bf M}$.

The macroscopic metric  geometry is obtained by an appropriate
spacetime averaging of the microgeometry; thus part of the
definition  of a cosmological model consists of specifying the
averaging scheme (which must be consistent with the physical
assumptions of the model encapsulated in the conditions C3 and C4
below) and the cosmological scale over which averaging or the
smoothing occurs (i.e., we must specify the averaging scale $\ell$
or averaging region). In particular, a given microgeometry may be
inhomogeneous on both small scales $\ell_s$ and large scales
$\ell_l$. The averaging scale $\ell$ will satisfy $\ell_s <<
\ell$, but it is possible that $\ell << \ell_l$, so that the
macrogeometry is still inhomogeneous (but `smooth') on large
scales.  As noted earlier, the scales for which the cosmological
model is defined must be explicitly specified.

\subsection*{C2. Timelike Congruence:} There exists a timelike
congruence $({\bf u})$ (in principle locally, but by definition we
can  extend this to the whole manifold), representing a family of
fundamental observers.

This congruence is associated with the 4-velocity of the averaged
matter in the model; i.e., the matter admits a formulation in
terms of an averaged matter content which defines an average
(macroscopic) timelike congruence. If there is more than one
matter component giving rise to more than one macroscopic timelike
congruence we identify a fundamental macroscopic timelike
congruence. There is always one, which has a physical meaning.
This leads to a covariant $3+1$ split of spacetime \cite{Ellis}.
Mathematically this means that the spacetime is
\emph{$\mathcal{I}$-non-degenerate} and hence the spacetime is
uniquely characterized by its scalar curvature invariants.

In addition to the formal parts C1 and C2 of the definition of a
cosmological model $({\bf M,g, \ell, u})$, we must also specify
the physical relationship (interaction) between the macroscopic
geometry and the matter fields, including how the matter responds
to the macroscopic geometry.

\subsection*{C3. Macroscopic field equations:} There exists an
appropriate set of macroscopic field equations relating the
averaged matter and appropriately averaged (or macroscopic)
geometry. This is based on an underlying microscopic theory of
gravity (such as, for example, GR), and an appropriate formalism
to average the geometry and find corrections (correlations) due to
averaging to the Einstein tensor in the resulting field equations:

\begin{equation}
\tilde{G}^{a}_{~b} + C^{a}_{~b}=T^{a}_{~b},
\end{equation}
in the usual geometrized units in which $c = 1 = {8\pi G}/{c^2}$,
where $\tilde{G}^{a}_{~b} \equiv \tilde{R}^{a}_{~b} -
\frac{1}{2}{\delta}^{a}_{~b}\tilde{R}$ and $\tilde{R}^{a}_{~b}$ is
the Ricci tensor of the averaged macrogeometry, $C^{a}_{~b}$ is
the correlation tensor, and $T^{a}_{~b}$ is the energy momentum
tensor (already assumed averaged).

This is the emphasis of the analysis here. Note that in this
context only the Ricci tensor needs to be averaged.

\noindent{\em Discussion}

The energy-momentum tensor that appears in the EFE (defined
formally by the variation of the matter action with respect to the
metric) depends on the gravitational field. Indeed, in GR the
gravitational energy contributes to the total energy-momentum. So
in this sense some aspects of the averaged gravitational field are
already included in the model. {\footnote{ To minimize this
effect, the mixed components of $T^{a}_{~b}$ are usually
considered.}}

In general, it does not follow from the contracted Bianchi
identities that the energy-momentum is conserved with respect to
the macrometric: $T^{c}_{~b;c} \neq 0$. Is it possible to define a
new effective  energy-momentum tensor $\hat{T}^{a}_{~b}$ which is
conserved? For example, $\hat{T}^{a}_{~b} \equiv
\tilde{G}^{a}_{~b} = T^{a}_{~b} - C^{a}_{~b}$ (which satifies the
macro-conservation equations with respect to the macrogeometry by
definition). However, it must be ensured that the relationship
between the effective matter and the (averaged) macrogeometry is
consistent with the underlying microphysical model and the
averaging scheme. At a heuristic level, this could be (within the
cosmological model under consideration) modelled qualitatively
through an appropriate (new type of) effective equation of state.

In particular, the separation between the gravitational field and
the matter is not scale invariant. Indeed, the notion of a perfect
fluid is not scale invariant \cite{LL}, and in the presence of a
gravitational field the equation of state of the matter is
consequnetly in general scale dependent \cite{Deb}. Thus,
averaging affects the effective equation of state. In the simplest
cosmological models we can write $T^{a}_{~b}= diag[-\rho,p,p,p]$
and $C^{a}_{~b} \equiv diag[-\rho_c,p_c,p_c,p_c]$, and we have that
$\hat{T}^{a}_{~b} = diag[\rho-\rho_c,p-p_c,p-p_c,p-p_c]$. A
physical equation of state relates $\rho$ and $p$, whereas the
effective equation state is a relationship between $\hat{\rho}
\equiv\rho-\rho_c$ and $\hat{p} \equiv p-p_c$. {\footnote{ If we
split up $\hat{T}$ into a standard matter part plus an
`additional' matter term, in general there exists an interaction
between them (i.e., each matter term is not necessarily conserved
\cite{Billyard}).}}

In this heuristic framework all of the effects of averaging go
into the redefined energy-momentum tensor $\hat{T}^{a}_{~b}$ and
the effective equation state of the macro-matter (subject to any
changes in the equations of motion of the macro-matter, although
we again note that $\hat{T}^{a}_{~b}$ is conserved relative to the
macrogeometery). Therefore, in this reinterpretation, any
non-standard equation of state is not due to any exotic matter,
but is the effect of averaging (cf. \cite{Bsph}). In particular,
an equation of state renormalization at the level of about 1\%
$\approx {\cal O}(\ell/\ell_H)$ naturally fits into this new
averaging interpretation (and especially an additional
$C^{a}_{~b}$ corresponding to spatial curvature arising from
averaging \cite{coley}).

\subsection*{C4. Equations of motion:} We also need to know the
trajectories along which the cosmological matter moves (and also
the light trajectories, which determine observational relations).
{\footnote{ This may add additional geometrical quantities (other
than the curvature) that need to be considered (and need to be
averaged). Perhaps this can be done using appropriate scalars, but
this is the subject of future investigation.}} In principle,
averaging the 4-velocities of the microscopic matter particles
does not necessarily give the macroscopic 4-velocity of the
cosmological matter {\footnote{ In GR a free point particle moves
on a timelike or null geodesic. But a system may behave as point
particles on small scales but {\em not} on larger scales
(alternatively, the definition of a point particle is also {\em
not} scale invariant). Therefore, macroscopic (averaged) matter
need not move on geodesics of the macrogeometry.}} (and the
average motion of a photon need not be a null geodesic in the
averaged geometry).

\noindent{\em Discussion}

This needs to be consistent with the averaged field equations: if
the matter satisfies the conservation equations in the macroscopic
geometry (which follows from the averaged field equations), then
the equations of motion must be consistent with this (i.e.,
cosmological dust follows macroscopic timelike geodesics). Note
that $\tilde{G}^{a}_{~b}$ satisfies the contracted Bianchi
identities (with respect to the macrogeometry). {\footnote{ Since
$\hat{T}$ is conserved the motion of matter is restricted; we
still need a model for the  motion of photons.}}

The fundamental congruence is essentially the average of the
timelike congruences along which particles move in the
microgeometry. This implies a relationship between matter and
macrogeometry. If the particles move on timelike geodesics in the
microgeomerty, does the fundamental congruence consist of timelike
geodesics (of the macrogeometry). Does (a beam of) photons move on
null geodesics of the macrogeometry. This raises the question of
whether it is possible to perform an appropriate averaging so that
on average the macroscopic matter moves on timelike/null geodesics
of the macrogeometry.

Ultimately we need a theory for the propagation of photons. The
motion of light in the macroscopic geometry must be consistent
with the limiting motion of timelike particles in the
macrogeometry. Alternatively, we could average the
Einstein-Maxwell equations and take the geometric optics limit
\cite{Dyer}; i.e., we assume that the actual photons satisfy the
Einstein-Maxwell equations in the optical limit, and thus follows
a null geodesic in the macrogeometry. We must then average the
Einstein-Maxwell equations \cite{Deb} and obtain a suitable form
for the macroscopic Einstein-Maxwell equations and an appropriate
(corresponding) optical limit in the macroscopic regime.

\subsection*{C5. Observations:} Finally, we need to be able to
relate averaged quantities with physical observables, which
ultimately must be consistent with cosmological data.

\subsection{Covariance}

We have advocated an approach to averaging a cosmological model
utilizing scalar invariants. Thus far we have focussed attention
on averaging the EFE (e.g., the Ricci tensor). We shall briefly
address the more general question of averaging or smoothing a
differentiable manifold in the next section. However, the second
aspect of averaging in a cosmological model involves the equations
of motion of matter and the geodesic structure.

Since there exists a global timelike congruence ${\bf u}$, global
covariance is broken (and the spacetime admits of a $1+3$ split).
This timelike congruence is defined physically, and in principle
is measurable. The resulting theory need only be invariant with
respect to restricted covariance \cite{EMa}. Perhaps, for
describing the equations of motion of particles and null rays, we
only need to consider scalars invariant with respect to restricted
sets of coordinate transformations (this would correspond to the
kinematic variables such as the shear and expansion scalars
associated with the macroscopic covariant derivative of ${\bf u}$
\cite{Ellis}).

This suggests that a description of the geodesic structure might
be possible utilizing averaged the restricted (kinematical) scalar
quantities. We shall not pursue this further here (cf.
\cite{buch}).

\subsection{Example: FLRW models}

In the standard FLRW model there are a number of simplifications
and assumptions. However, the cosmological model is not fully
defined in the sense above and hence has a lack of physical
predictability.

The past approaches to averaging have been ideally suited to the
FLRW models (with small, vanishing in the limit, perturbations).
Therefore, an extension of these approaches to cosmological models
other than FLRW models is not possible \cite{Deb}. Also, some of
these approaches deal exclusively with dust. Therefore, in these
approaches to averaging and cosmology it is impossible to obtain
the potentially most important effects.

Let us examine the assumptions C1 -- C5 in the usual FLRW plus
perturbations model. The macrometric ${\bf g}$ is the FLRW metric
(C1) and ${\bf u}$ also has a geometric meaning (C2). In the usual
point of view there are no correlations due to averaging (i.e.,
the correlation tensor is zero; $C^{a}_{~b}=0$) or, more
precisely, they are negligible (C3). In this case it follows from
the contracted Bianchi identities that energy-momentum is
conserved: $T^{c}_{~b;c}=0$, which relates the matter to the
averaged geometry. All other effects are assumed negligible (C4).

However, there is no formal proof that such assumptions arise from
a rigorous averaging scheme of some appropriate (physically
motivated) microgeometry. In addition, there are some important
effects which are not necessarily small perturbations. In
\cite{coley} it was argued that important effects at 1\% level
cannot be neglected.

Since there is no scale included in the model, in a sense the
model is incomplete. Indeed, the model does not even have the
ability to determine whether there is a scale above which the
geometry is exactly FLRW or whether at all scales the geometry is
only approximately FLRW (with a given perturbation scale). Indeed,
in cosmological perturbation theory, both the scale of the
background and the scale of perturbations need to be specified,
neither of which are given since no notion of order of
approximation is included \cite{Ellis,BuchEllElst,BuchEll}. In
addition, the motion of photons must be independently postulated.

Regarding C5 we ask the question: does this model agree with
observations? If it does not, then even if the model agrees in
some approximate sense with most observations, there is no
structure within which to discuss the potential small
discrepancies with observed data, which is a deficiency of the
model. If the model does, then it would be remarkable, although
there is still the need for a physical explanation of the dark
energy.

In the standard approach, there might be a spatial curvature term
on the left-hand side of the macroscopic field equations with
intrinsic curvature parameter $k$, and an effective spatial
curvature term on the right-hand side of the macroscopic field
equations (due to an effective equation of state arising from
averaging) with curvature parameter $\hat{k}$ (which could be
thought of as a `renormalization' to the intrinsic curvature
parameter). If observations indicate that $\Omega_k \approx 1-2\%$
\cite{Riess:2006fw,bennett}, then there is no physical mechanism
within the model (particularly if there is an inflationary period)
to produce an intrinsic curvature parameter $k$ of this magnitude,
whereas an effective curvature parameter $\hat{k}$ of about a
percent arises naturally from averaging \cite{Col}.

\newpage

\section{An approach to averaging cosmological models using scalar invariants}

Since for a cosmological model the geometry is completely
characterized by its scalar curvature invariants \cite{NEWINV}, we
shall adopt the approach that we shall only average the scalar
curvature invariants (thereby avoiding the problems mentioned
earlier). Therefore, we have a microgeometry completely
characterized by its set of scalar curvature invariants
$\mathcal{I}$. We then average these microgeometry scalar
curvature invariants to obtain a new set of macrogeometry scalar
curvature invariants  $\tilde{\mathcal{I}}$, which now completely
characterizes the macrogeometry. In general, this macrogeometry is
unique (for a given averaging region). Let us consider first the
more general mathematical question of averaging the geometry; we
shall then focus attention on the physical case of a cosmological
model.

\subsection{Averaging the geometry}

In the general mathematical context we want to describe the
averaged geometry (represented by the Riemann tensor and its
covariant derivatives) and interpret the results. Let us consider,
$\mathcal{I}\equiv\{R,R^{\mu}_{~\nu}R^{\nu}_{~\mu},\dots\}$, which
is an ordered set of functions on {\bf M}. $\tilde{\mathcal{I}}$
is then also an ordered set of functions. If we write
$\tilde{\mathcal{I}}\equiv\{\tilde{R},\widetilde{R^{\mu}_{~\nu}R^{\nu}_{~\mu}},\dots\}$,
then of course it does not follow that, for example,
$\widetilde{R^{\mu}_{~\nu}R^{\nu}_{~\mu}}=\tilde{R}^{\mu}_{~\nu}\tilde{R}^{\nu}_{~\mu}$
(under averaging). But the question is: does the ordered set of
functions $\tilde{\mathcal{I}}$ correspond to the associated
scalar curvature invariants for some metric $\tilde{g}$ (which
will then serve to {\em define} the macrometric $\tilde{g}$). This
raises some interesting mathematical questions (see below). In
principle, this is not necessarily true in general.

Indeed, we note that it is not generally true that the set of
averaged scalar invariants $\tilde{\mathcal{I}}$ actually
determines a Riemannian geometry. In addition, if $A=B$, then
$\bra{A}\ket=\bra{B}\ket$. 
{\footnote{ For convenience we use the  simpler notation $\bra{A}\ket \equiv
\tilde{A}$ here and in the last section.}} If $A^2=B^2$, then
$\bra{A^2}\ket=\bra{B^2}\ket$, 
but this does not imply
$\bra{A}\ket^2=\bra{B}\ket^2$. In this particular case it is true
because if $A,B$ are scalars, then $A^2=B^2$ implies $A= \pm B$
and $\bra{A}\ket= \pm \bra{B}\ket$; but if (as in the static
example below) $(R2)^2=(R1)^3$, while it is true that
$\bra{(R2)^2}\ket=\bra{(R1)^3}\ket$, it does not follow in general
that $\bra{R2}\ket^2=\bra{R1}\ket^3$. However, relationships
of this form will be true
for many of the applications of interest (see the next section).

Indeed, it is plausible that the ordered set of functions
$\tilde{\mathcal{I}}$ correspond to the associated scalar
curvature invariants for some macrometric $\tilde{g}$ for the
class of $\mathcal{I}$-non-degenerate geometries that constitute
the class of cosmological models defined here. Since the
geometries are $\mathcal{I}$-non-degenerate and in 4D properties
of the geometry can be represented in terms of scalars, and since
relations between different terms (functions) in the set
$\mathcal{I}$ (e.g., $R$ and $R^2$ are functionally dependent) and
the corresponding terms in the set $\tilde{\mathcal{I}}$ (e.g.,
$\tilde{R}$ and $\tilde{R}^2$) are functionally related in exactly
the same way and syzygies (e.g., describing the algebraic type)
are maintained under averaging, it follows that in general the set
$\tilde{\mathcal{I}}$ uniquely gives rise to a macrometric
$\tilde{g}$ (which will have similar algebraic properties to the
micrometric ${g}$). {\footnote{ This has not been proven here, but
it is plausible; if there are any exotic counterexamples, then the
definition of a cosmological model, and the rules of averaging,
could be suitably amended to avoid this. It is true for the
cosmological models of physical interest. An alternative approach is suggested
in the scalar averaging procedure described below.}}

We note that a subset of 4D geometries are uniquely determined by
their curvature only (i.e., no covariant derivatives are
necessary), so that we only need to consider the scalar polynomial
invariants constructed from the curvature tensor (i.e., an
appropriate subset of $\tilde{\mathcal{I}}$) {\footnote{ That is,
we do not need to consider invariants like 
$di\mbox{Weyl} \equiv C_{abcd;e} C^{abcd;e}$ or $di\mbox{Ricci} =
R_{ab;e} R^{ab;e}$,
which depend on derivatives of the metric greater than 2, which
may add new problems regarding averaging.}} For algebraically
general geometries this is indeed true, and it is certainly true
for many of the important applications we are interested in. In
addition, as noted above, we are primarily interested in the Ricci
curvature (and the 4 Ricci scalar invariants) in applications in
cosmology.

If a spacetime is not $\mathcal{I}$-non-degenerate it can not be
uniquely determined by its scalar polynomial invariants. There is
one subclass of $\mathcal{I}$-non-degenerate spacetimes which, in
a certain sense, can be determined by its scalar polynomial
invariants; namely the type $D^k$ spacetimes (i.e., spacetimes for
which the curvature and all of its covariant derivatives are
simultaneously of algebraic type D) \cite{NEWINV}; this subclass
includes the static spherically symmetric spacetimes of particular
interest here.

\subsubsection{Uniqueness}

The macrogeometry is completely specified by the set
$\tilde{\mathcal{I}}$. Suppose that  $\tilde{\mathcal{I}}$ (or
some appropriate subset of  $\tilde{\mathcal{I}}$) is derivable
from a macrometric $\tilde{g}$. Although this macrogeometry is
unique, it is possible for two different (non-isomorphic)
microgeometries to give rise to the same macrogeometry. However,
the algebraic properties  (i.e., the algebraic properties of the
Riemann tensor and its covariant derivatives) of the underlying
microgeometry and the resulting macrogeometry must be the same.

Since in the cosmological models under consideration the
microgeometry is completely characterized by the set of scalar
curvature invariants $\mathcal{I}$, the algebraic properties of
the underlying microgeometry is generally determined through the
syzygies of $\mathcal{I}$. {\footnote{ A particular cosmological
spacetime (an exact solution or a subset of spacetimes with
arbitrary functions in 4D) is often completely determined by
syzygies involving scalar invariants.}} Any syzygy $\mathcal{S}$
is maintained under averaging (integration), giving rise to a
corresponding syzygy $\tilde{\mathcal{S}}$ of the set
$\tilde{\mathcal{I}}$ and then the averaged geometry will obey a
similar set of (averaged versions) of these syzygies; that is, the
averaged geometry will be of the same type as the microgeometry,
and hence the macrogeometry is at least as algebraically special
as its underlying microgeometry.{\footnote{In principle, new
syzygies may appear via averaging and so it is possible, in
principle, for the macrogeometry to be more algebraically special
than its underlying microgeometry.}} In addition, if certain terms
in $\mathcal{I}$ satisfy certain algebraic conditions due to the
differential Bianchi identities, then the corresponding terms in
the set $\tilde{\mathcal{I}}$ would satisfy a corresponding set of
algebraic conditions, consistent with the Bianchi identities of
the macrogeometry. Therefore, in applications in cosmology this
approach will produce a unique macrogeometry.

\subsubsection{Proposal: Scalar Averaging Procedure}

Let us consider the ordered set of functions
$\mathcal{I}$ on {\bf M} in the form:

$$\mathcal{I} \equiv \{R,R1,R2,R3, R^2,R^{\mu}_{~\nu}R^{\nu}_{~\mu},\dots, C^2, \dots\}.$$

First, let us omit any scalars from this set that are not algebraically independent
(e.g., $\{R^2,R^{\mu}_{~\nu}R^{\nu}_{~\mu}\dots\}$)
{\footnote{ Including, for example, the differential Bianchi identities.}} to obtain an
(the appropriate `independent') subset $\mathcal{I}_A$. Second, 
for a particular spacetime, we omit any scalars from $\mathcal{I}_A$
that can be obtained from syzygies defining that particular spacetime
(e.g., defining the algebraic type of the spacetime, such as the Segre type or 
the Petrov type). For example, for a Ricci tensor corresponding to the algebraic
form of a perfect fluid we could omit $\{R2,R3\}$ (relative to $\{R,R1\}$). We consequently obtain the
subset $\mathcal{I}_{SA}$: e.g.,

$$\mathcal{I}_{SA} \equiv \{R,R1,\dots, C^2,\dots\}.$$
For the spacetimes under consideration the
microgeometry is then completely characterized by the (sub)set of scalar
curvature invariants $\mathcal{I}_{SA}$. {\footnote{ This subset is not unique;
perhaps a scheme should be adopted whereby higher order scalars should always be omitted;
e.g., $R^2$ should be omitted relative to $R$.}}

We now construct the new ordered set of functions $\tilde{\mathcal{I}}_{SA}$ by averaging
the various scalar invariants of $\mathcal{I}_{SA}$:

$$\tilde{\mathcal{I}}_{SA}\equiv\{\tilde{R}, \widetilde{R1}, \dots, \widetilde{C^2},\dots\}$$
where all of the original scalar invariants omitted from the original set $\mathcal{I}$
are replaced by a new set of functions obeying exactly the same algebraic properties
(or syzygies)  as $\mathcal{I}_{SA}$. Therefore, it is assumed that $\tilde{\mathcal{I}}_{SA}$
comes equiped with these syzygies, so that we could construct the corresponding set $\tilde{\mathcal{I}}$
consisting of the members of $\tilde{\mathcal{I}}_{SA}$ and all of the corresponding syzygies.
Consequently, the set $\mathcal{I}_{SA}$
is an ordered set of functions (scalar curvature invariants) on {\bf M} which 
uniquely determines the macrogeometry
with exactly the same algebraic properties as the original microgeometry.

We shall refer to this proposal to obtain the set $\mathcal{I}_{SA}$ and the
associated averaged (macro) geometry as the {\bf Scalar Averaging Procedure}.

Note that in the cosmological application, in which the Ricci
tensor is the fundemental object, the simple set  
$\tilde{\mathcal{I}}_{SA}\equiv\{\tilde{R}, \widetilde{R1}\}$
completely characterizes the averaged Ricci tensor (cf. \cite{buch}).
However,  this set does not completely determine the macrogeometry. In addition,
we still need to know
the trajectories along which the cosmological matter and null rays
move; that is,  macrogeometric effects not arising from the Ricci tensor
alone.

\subsection{Cosmological model}

In the case of a cosmological model, we only need to be able to
average the Ricci tensor (or Einstein tensor) that appears in the
EFE.

\subsubsection{Ricci tensor}

Since from C3 we have an effective set of field equations:
\begin{equation}
\tilde{R}^{a}_{~b}+C^{a}_{~b}=T^{a}_{~b},
\end{equation}
we only need to consider the macrogeometric Ricci tensor
$\tilde{R}^{a}_{~b}$ here (the correlation tensor is obtained from
the averaging procedure). The microgeometric Ricci tensor
${R}^{a}_{~b}$ is completely characterized by a set of scalar
curvature invariants $\mathcal{I}_R$. Averaging these scalar
curvature invariants we obtain the set
$\tilde{\mathcal{I}}_{\tilde{R}}$, which completely characterizes
the macrogeometric Ricci tensor $\tilde{R}^{a}_{~b}$. Since
constructing the Ricci tensor from a set of scalar curvature
invariants $\mathcal{I}_R$ is relatively simple compared to the
corresponding problem for the Riemann tensor, and since the
reduced set of scalar curvature invariants $\mathcal{I}_R$ is
considerably smaller than $\mathcal{I}$, we have considerably
reduced the complexity of the problem in this new averaging
approach. Indeed, for a Ricci tensor of the algebraic form of a
perfect fluid, there are effectively (only) two independent scalar
invariants, the Ricci scalar and a single Ricci eigenvalue
(corresponding to the effective energy density, $\rho$, and
pressure, $p$, of the perfect fluid). Therefore, in the context
of the scalar averaging procedure, we have the set
$\{\tilde{R}, \widetilde{R1}\}$.

We also note that the syzygies of the macrogeometry Ricci tensor
must be consistent with the syzygies of the energy-momentum tensor
through the macroscopic field equation. If the energy-momentum
tensor tensor is of the form of a perfect fluid; i.e.,
${T}^{a}_{~b} = diag[-\rho,p,p,p]$, then ${T}^{a}_{~b}$ obeys a
number of syzygies. A relation (equation of state) between $\rho$
and $p$ would be represented by a further algebraic syzygy. For
example, defining ${A}^{a}_{~b} = {T}^{a}_{~b} - \frac{1}{2} T
{\delta}^{a}_{~b}$, where $T$ is the trace of ${T}^{a}_{~b}$
(compare with the form of the Ricci tensor in the EFE), the
equation of state $\rho+3p=0$ corresponds to the syzygy
$A^2=\frac{1}{3}{A}^{a}_{~b}{A}^{b}_{~a}$, where $A$ is the trace
of ${A}^{a}_{~b}$. {\footnote{ Since this particular syzygy can
only be satisfied if $\rho+3p=0$ or $5\rho+3p=0$, if the
cosmological matter satisfies appropriate energy conditions, this
syzygy uniquely specifies the equation of state $\rho+3p=0$ .}}
{\footnote{ Therefore, note that in order to obtain a macroscopic
`spatial curvature' it is necessary to include a cosmological
pressure.}}

Finally, we note that if every member of $\mathcal{I}_R$ is zero
(which uniquely characterizes a Ricci flat spacetime), then every
member of ${\tilde{\mathcal{I}}}_{\tilde{R}}$ is also zero, and
the macrogeometric Ricci tensor is also zero (flat). Therefore it
follows that the average of a microvacuum ${R}^{a}_{~b}=0$ gives
rise to a macrovacuum ${\tilde{R}}^{a}_{~b}=0$ (and the
corresponding correlation tensor is zero, and the macroscopic
field equations are trivial). Therefore, within this approach, no
new effects due to averaging occur at the level of the macroscopic
field equations when there is no cosmological (average) matter
content (i.e., a non zero energy momentum tensor ${T}^{a}_{~b}$ is
needed). {\footnote{ Note that in this approach, if the
microgeometry is Ricci flat then the macrogeometry is Ricci flat.
In vacuum, there will be no local variations (local
inhomogeneities) in the Ricci tensor since there is local source
with local inhomogeneities with a physical intrinsic scale.
Indeed, the only inhomogeneities in the geometry must therefore
come from (not from the energy-momentum tensor through the EFE
but) physical boundary conditions, which would consequently be
large scale inhomogeneities. Therefore, there would be no physical
mechanism to introduce local inhomogeneities whose average would
lead to a non-Ricci flat macrogeometry.}} As we noted above, the
primary focus of this paper is C3 and the effective macroscopic
field equations.

However, this does not imply that there are no averaging affects
in the case of vacuum. Since from C4 we also need to be able to
relate averaged quantities with observables and we need to know
the trajectories along which the cosmological matter and null rays
move, macrogeometric effects (not arising from the Ricci tensor
alone) will indeed play a role. However, this is beyond the scope
of the present work. It may be possible to express the effects in terms of kinematic
scalars which are invariant with respect to all coordinate changes
that preserve the fundamental congruence.

\subsubsection{Interpretation}

Finally, it is necessary to determine whether the correlations due
to averaging alter the geometry or affect the effective
energy-momentum tensor. This is partly a question of
interpretation. This must be done within the context of the
underlying cosmological model. We shall address this in the simple
example in the next section.

In particular, in the cosmological application it may be
appropriate to reinterpret the averaging correlations as
corrections to the matter fields through the EFE. Thus,
concentrating on the Ricci tensor again, writing
$\tilde{G}^{a}_{~b}=T^{a}_{~b}$ and defining
${\cal{G}}^{a}_{~b}=$$\tilde{G}^{a}_{~b} - C^{a}_{~b}$, we can
rewrite this as ${\cal{G}}^{a}_{~b}={\cal{T}}^{a}_{~b}$ (by
absorbing the correlation tensor $C^{a}_{~b}$ into
${\cal{T}}^{a}_{~b}$). We can now demand that the new macro-Ricci
tensor ${\cal{R}}^{a}_{~b}$ (corresponding to
${\cal{G}}^{a}_{~b}$) has exactly the same algebraic properties as
the Ricci tensor of the microgeomety, so that the averaging
correlations are interpreted in the cosmological context as
appropriate corrections to the matter fields. That is, we have
defined the new averaged Ricci tensor ${\cal{R}}^{a}_{~b}$,
derivable from some appropriate averaged geometry with an
appropriate micro-metric, and some of the correction terms have
been included in the new effective energy-momentum tensor
${\cal{T}}^{a}_{~b}$.

\newpage

\section{Example: Static Spherically Symmetric  Perfect Fluid Spacetimes}

We shall consider the specific example of a static spherically
symmetric perfect fluid spacetime in this section. This is an
appropriate simple model for illustration since it can include an
arbitrary function of one variable, there is a non-vanishing
pressure, the averaging region does not change with time and there
are no gravitational waves.

Let us write the static spherically symmetric perfect fluid line element
in the form
\begin{equation}
\label{eqn4.4}
ds^2 = - e^{f(r)} dt^2 + e^{g(r)} [dr^2 + r^2 (d\theta^2 + \sin^2
\theta d \phi^2)].
\end{equation}
The two arbitrary functions $f(r)$ and $g(r)$ satisfy the
differential constraint
\begin{equation}
\label{eqn4.5}
\frac{d^2g}{dr^2} - \frac{1}{r} \frac{dg}{dr} - \frac{1}{r}
\frac{df}{dr} - \frac{df}{dr}\frac{dg}{dr} - \frac{1}{2}
\left(\frac{dg}{dr}\right)^2
+ \frac{1}{2} \left(\frac{df}{dr}\right)^2 + \frac{d^2f}{dr^2} = 0,
\end{equation}
in order for the Einstein tensor to be of the form of a
perfect fluid.

The 16 CM polynomial scalar curvature invariants \cite{CM}, as
given in GRTensor \cite{grtensor}, for the static spherically
symmetric spacetime are then as follows (after the constraint has
been applied). The 4 (linear, quadratic, cubic and quartic) Ricci
tensor scalar invariants (which are related to the Ricci scalar
and the three Ricci tensor eigenvalues) are:
\begin{equation}
\label{eqn4.6}
R = \frac{1}{2} e^{-g} \left[ 2 \frac{d^2f}{dr^2} - 5 \left(\frac{df}{dr}
\right)
\left(\frac{dg}{dr} \right) +
\left(\frac{df}{dr} \right)^2 -3\left(\frac{dg}{dr} \right)^2
- \frac{8}{r} \frac{df}{dr} - \frac{12}{r} \frac{dg}{dr}\right]
\end{equation}

\begin{equation}
\label{eqn4.7}
R1 = \frac{3}{64} e^{-2g}  \left[2 \frac{d^2f}{dr^2} - \left(\frac{df}{dr}
\right)
\left(\frac{dg}{dr}\right) - \left(\frac{dg}{dr}\right)^2 +
\left(\frac{df}{dr}\right)^2
- \frac{4}{r} \frac{dg}{dr}  \right]^2
\end{equation}
and $R2$ and $R3$ are proportional to $(R1)^{3/2}$ and $(R1)^2$,
respectively.

The $4$ polynomial Weyl scalar invariants are:
\begin{equation}
\label{eqn4.8}
W1R = \frac{1}{24} e^{-2g} \left[ 2 \frac{d^2f}{dr^2} + \left(\frac{df}{dr}
\right)^2
- 2 \left( \frac{df}{dr} \right)\left(\frac{dg}{dr}  \right) - \frac{2}{r}
\frac{df}{dr} \right]^2
\end{equation}
$W2R \propto (W1R)^{3/2}$ and $W1I = W2I = 0$, and the mixed
invariants are $M3= M2R \propto (R1) (W1R)$, $M4 \propto (R1)
(W1R)^{3/2}$, $W5R \propto (R1)^{3/2} (W1R)$ and $M1R = M1I = M2I
= M5I = 0$. Note that GRTensor is also able to compute the
differential scalar invariants
 $di\mbox{Weyl}$ and $di\mbox{Ricci}$.

 In the case of a static spherically symmetric constant curvature spacetime
 with $f(r) =1$, $g(r) = (1+kr^2)^{-2}$ ($k$ constant), we have that:
\begin{equation}
\label{eqn4.9}
R = 24k, \quad R1 = 12k^2
\end{equation}
$(R2=24k^3$, $R3 = 84k^2)$, and all other scalar invariants (including the
differential invariants) vanish.

In the case of the static Schwartzchild-deSitter
(Kottler \cite{kramer}) spacetime with $f(r) =1 - \frac{2m}{r}
- \frac{\Lambda r^2}{3} = [g(r)]^{-1}$, we have that:
\begin{equation}
\label{eqn4.10}
 R=4 \Lambda, W1R = \frac{6m^2}{r^6}, W2R = \frac{-6m^3}{r^9},
di\mbox{Weyl} = -240(6m -3r + \Lambda r^3)\frac{m^2}{r^9}
\end{equation}
and all other scalar invariants (including diRicci) are zero.  For
the Schwarzchild vacuum solution $\Lambda =0$.  In the case of an Einstein space
(with $m=0)$, the only non-vanishing scalar invariant is $R = 4\Lambda$.

The approach outlined in the previous section is then as follows.
Averaging the scalar invariants we obtain, for example, $\bra R \ket$
and $\bra R1\ket$ ($\bra W1R \ket$ and so on, averaging
 eqns. (\ref{eqn4.6})-(\ref{eqn4.8}), where $\bra{R}\ket \equiv \tilde{R}$)
 together with an appropriately averaged constraint eqn. (\ref{eqn4.5}).
 These are then the scalar curvature invariants of the averaged static
 spherically symmetric geometry. As noted above, we will focus on $\bra R \ket$
and $\bra R1\ket$ here.

 \subsection{Averaging Scales}

 Suppose that the averaging scale is $\ell \equiv L$.  Suppose also
 that there are inhomogeneities with scales $\beta^{-1}$ and $\lambda^{-1}$,
 where $\beta^{-1}$ of order unity ($\beta \sim 1$, $L < \beta^{-1}$;
 i.e., global inhomogeneities) and $\lambda^{-1}$ is a small scale
 $(\lambda^{-1} <<1$, $\lambda = \frac{1}{\alpha} L$ where $\alpha$ is
 large (integer) so that $\lambda/L < 1$; i.e., local small scale
 inhomogeneities).
 Hence, the function $f$ varies slowly with respect to $\beta r$, and varies
 quickly with respect to $\lambda r$ (on smaller scales).

 Let us write the scale dependence explicitly as
 \begin{equation}
\label{eqn4.11}
f = f (\beta r, \lambda r),
\end{equation}
where we effectively treat $\lambda r$ as a separate variable.
Consider $r = r_0$, and a neighbourhood of $r_0$, $I \equiv (r_0 -
\frac{L}{2}, r_0 + \frac{L}{2})$ of length $L$ (the averaging
region).  Define $r_0 + r^\prime \in I$, so that $r^\prime \in
(-\frac{L}{2}, \frac{L}{2})$ which parameterizes points in $I$. We
can write (set $\beta =1)$
\begin{equation}
\label{eqn4.12}
f(r) = f(r_0, \lambda(r_0 + r^\prime))
\end{equation}
in $I$, where $f(r_0) = f(r_0, \lambda r_0)$.  Therefore, in
$I$, $f(r)$ has small scale variations with respect to $\lambda r^\prime$.
We can write
\begin{equation}
\label{eqn4.13}
f(r) \simeq f(r_0, \lambda r_0) + {\cal O} (\mu)
\end{equation}
where $\mu$ is the small scale of the amplitude of inhomogeneities
in $I$ ($\mu << 1$).

We now average $f(r)$ over these small scale
inhomogeneities in $I$, and define the average $\bra f(r_0) \ket
= \bra f(r, \lambda r) \ket |_{r=r_0}$ by:
\begin{equation}
\bra f(r_0) \ket = \frac{1}{L} \int^{L/2}_{-L/2} f(r_0,
\lambda (r_0 + r^\prime)
dr^\prime
\end{equation}
(i.e., we are effectively `averaging over $\lambda r$').
Notice that there
are a number of scales in the problem: $\frac{1}{\lambda} << 1$, $\mu << 1$,
$\frac{1}{\alpha} = \frac{\lambda}{L} < 1$.  Note also that
\begin{equation}
\frac{\partial f}{\partial r} = \frac{\partial}{\partial (\beta r)}
f(\beta r, \lambda r) + \frac{\partial}{\partial (\lambda r)}
 f(\beta r, \lambda r),
\end{equation}
where the second term is a dominant fast varying term.

We now Fourier analyse the functions $f(r)$ and $g(r)$ in the
averaging region $I$ with respect to $(\lambda r^\prime)
~[{\lambda}^{-1} = {\alpha}/{L}]$:
\begin{eqnarray}
f(r) & \equiv & f(r_0) + \mu f_1(r_0) \sin \left(\frac{\alpha \pi r^\prime}
{L} \right) + \sum_{n=2} \mu^n f_n (r_0) \sin \left( \frac{n
\alpha \pi r^\prime}{L} \right) \nonumber \\
g(r) & \equiv & g(r_0) + \mu g_1(r_0) \sin \left(\frac{\alpha \pi
r^\prime} {L} \right) + \sum_{n=2} \mu^n g_n (r_0) \sin \left(
\frac{n \alpha \pi r^\prime}{L} \right)
\end{eqnarray}
where $f_1(r_0)$, $f_n(r_0)$, $g_1(r_0)$, $g_n(r_0)$ are slowly
varying functions of $r$ (i.e., their derivatives are small). Note
that
\begin{eqnarray}
\bra f(r_0) \ket & = & f(r_0)  \nonumber\\
\bra g(r_0) \ket & = & g(r_0).
\end{eqnarray}

Calculating, we obtain (with an abuse of notation)
\begin{eqnarray}
\label{eqn4.18}
g_r(r) & = & g_r(r_0) + \frac{\alpha \mu}{L} \pi g_1 \cos
\left(\frac{\alpha \pi r^\prime}{L}\right) + \frac{\mu^2 \alpha}{L}
 2 \pi g_2 \cos \left( \frac{2 \alpha \pi r^\prime}{L}  \right)  \nonumber\\
 && + \mu g^\prime_1 \sin \left(  \frac{\alpha \pi r^\prime}{L}
 \right) + \mu^2 g^\prime_2 \sin \left( \frac{2 \alpha \pi r^\prime}{L}
  \right) + {\cal O} (\mu^3) + {\cal O} \left( \frac{\mu^2 \alpha^2}{L^2}
   \right) \nonumber\\
   g_{rr}(r) & = & g_{rr} (r_0) - \frac{\mu\alpha^2}{L^2} \pi g_1 \sin
\left(\frac{\alpha \pi r^\prime}{L}\right)- \frac{\mu^2 \alpha^2}{L^2}
 4 \pi g_2  \sin \left( \frac{2 \alpha \pi r^\prime}{L}  \right)
  \nonumber\\
 && + {\cal O}(\mu^3) + {\cal O}   \left(  \frac{\mu\alpha^3}{L^3}
   \right)
\end{eqnarray}
We can now average the constraint equation (\ref{eqn4.5})  [using
various trigonometric formulae in evaluating the integrals]: To
zeroth order we obtain the constraint in terms of $f(r_0)$ and
$g(r_0)$ [effectively eqn. (\ref{eqn4.5})].  To next leading order
we obtain
\begin{equation}
\label{eqn4.19}
f_1 g_1 + \frac{1}{2} g^2_1 - \frac{1}{2} f^2_1 = 0
\end{equation}
(and to higher orders:  $f^\prime_1 g^\prime_1 + \frac{1}{2}
(g^\prime_1)^2 - \frac{1}{2} (f^\prime_1)^2 = 0$, $f_2 g_2 +
\frac{1}{2} (g^2_2) - \frac{1}{2} f^2_2 = 0$, etc.).  For example,
using the first of eqns. (\ref{eqn4.18}) to expand $g_r(r)$ and
$f_r(r)$ and integrating we obtain
\begin{eqnarray}
\bra \frac{df}{dr} \frac{dg}{dr} \ket & = & f_r(r_0) g_r(r_0) +
\frac{\pi^2}{2} f_1g_1 \left(\frac{\alpha}{L}\right)^2
\mu^2 + \frac{1}{2}
f^\prime_1 g^\prime_1 \mu^2 \nonumber\\
&& + \frac{\pi^2}{2} f_2 g_2 \left(\frac{\alpha}{L}\right)^2
\mu^4 + \frac{1}{2}
f^\prime_2 g^\prime_2 \mu^4 + {\cal O} \left(\frac{\alpha^4 \mu^2}{L^4}\right)
+ {\cal O} (\mu^6),
\end{eqnarray}
Therefore, by a direct calculation (and using the constraint
(\ref{eqn4.19}), we find that
 \begin{equation}
 \label{eqn4.21}
 \bra R\ket = R(r_0) \left(1 + \frac{1}{4L} g_1 \mu^2\right) -
 \frac{3 \pi^2}{4}
 e^{-g(r_0)}  \left(\frac{\alpha}{L}\right)^2 \mu^2
 \frac{1}{L} \left(f_1 g_1 + {\cal O} \left(\frac{1}{L}\right)\right)
 + {\cal O} (\mu^4)
\end{equation}
We also find that
\begin{equation}
\label{eqn4.22}
\bra R1 \ket = R1 (r_0) \left\{ 1 + {\cal O} \left(\frac{\mu^2}{L}
\right) \right\} + \frac{64 \pi^2}{3} F(r_0) \left( \frac{\alpha}{L}
 \right)^2 \mu^2 \frac{1}{L} (f_1 g_1),
\end{equation}
where $F(r_0)$ is a specific function of $r_0$ (not explicitly displayed
here).

\subsection{Interpretation} The ${\cal O} (\frac{L}{\mu^2})$
correction in the first term of $\bra R\ket$ in eqn.
(\ref{eqn4.21}) is a higher order renormalization term. To lowest
order we have that
$$ \bra R \ket = R(r_0) - \frac{3\pi^2}{4} e^{-g(r_0)}
\left(\frac{\alpha^2\mu^2}{L^3}\right)
(f_1 g_1)$$
and hence
\begin{eqnarray}
 \bra R\ket^2 & \cong & R^2 (r_0) - \frac{3 \pi^2}{2}
e^{-g(r_0)} R(r_0) \left(
\frac{\alpha^2\mu^2}{L^3}\right) \left(f_1 g_1\right)  \nonumber\\
& \equiv & R^2 (r_0) - \frac{3}{2}
G(r_0) \epsilon
\end{eqnarray}
where $G(r_0) \equiv e^{-g(r_0)} R(r_0)$ and $\epsilon \equiv
\frac{\pi^2\alpha^2 \mu^2}{L^3} (f_1 g_1) \simeq$ constant. Also,
we have that
\begin{equation}
\bra   R1\ket = R1 (r_0) + \frac{64}{3} H(r_0) \epsilon,
\end{equation}
where $H(r_0)= R1(r_0) F(r_0)$.  For a space of constant curvature
$k$ with $R(r_0) = 24k$, $R1(r_0) = 12 k^2 = \frac{1}{48}
(24k)^2$, we have that $G(r_0)= 768\gamma_0 k^2$, $H(r_0) =
-\frac{9}{16} k^2 \gamma_0$ (where $\gamma_0$ is a constant).
Therefore, we have
\begin{eqnarray}
\bra R\ket^2 & = & R^2(r_0) + 2(24k)^2 (-\gamma_0 \epsilon)\nonumber\\
\bra R1\ket & = & R1(r_0) + \frac{1}{48}(24k)^2 (-\gamma_0
\epsilon).
\end{eqnarray}
and we can interpret the average correlations as contributing a
small, ${\cal O}(\alpha^2 \mu^2 L^{-3})$, constant curvature term,
arising from the averaging of local inhomogeneities in the
micro-Ricci tensor, to the smooth macro-Ricci tensor (consistent
with the results of \cite{CPZ}).

\newpage

{\em Acknowledgements}. 
I would like to thank Sigbjorn Hervik and Robert van den Hoogen for helpful comments.
This work was supported by NSERC of Canada.

\end{document}